# UNDERSTANDING THE IMPACTS AND INFLUENCES OF METRO RAIL ON URBAN ENVIRONMENT –CASE STUDIES AND THE BENGALURU SCENERIO

R.Hemasree[1,*], C.V.Subramanian[2]

[1] Research Scholar, Periyar Maniammai Institute of Science and Technology, Thanjavur, India

[2] Professor, Periyar Maniammai Institute of Science and Technology, Thanjavur, India

*Corresponding Author: hemasree.raj@gmail.com

**Abstract:**

Urban sprawl is brought on by migration brought on by the new and exciting prospects in the cities. Cities expand rapidly, which leads to increased population growth and exceeding vehicle expansion. A city can increase its commercial prospects, reduce sprawl, and promote a feeling of community through transit-oriented development by incorporating public transportation options and considerations into more comprehensive economic and land use planning. Public transportation benefits both users and non-users since it decreases travel times, air pollution, energy use, and traffic congestion. To better understand the features and advantages of Metro Transit, surveys on car ownership, transit use frequency, and trip purpose are crucial. These surveys are nothing more than case studies of specific transit operators from all throughout the country. Case studies explore the various city conditions, the role of MRTS problems which are created by them, problems solved by them and to envisage the future problems over a period of time. They also provide the clarity of the various issues and challenges faced by the city at different time of the day. Bengaluru is an outstanding example how a pensioner's paradise has resulted now as a Meta city. It has undergone a huge metamorphosis trying to solve the various demands put forth by the economic activities. Transit oriented development-(TOD) seems the logical way to address the traffic and transportation problems which results in a mixed-use development, high rise, high density. After Delhi and Hyderabad, Bengaluru has entered the queue to embrace Namma Metro and implement Phases 2A and 2B after Phases 1 and 2. An inventory into selected case studies will relate and differentiate the conditions and systems in Bengaluru thus enriching our capability to tackle the Transit Oriented Development (TOD) issues at bay.

# Introduction:

As per report, "a second large urban rail revolution is spreading around the World,"(Kenworthy, 2015). Currently, nations in the US, Australia,, Canada, the Middle East, Europe, and Asia are experiencing a "trend back to urban rail." Urban rail is now being developed in both the conventional automobile-dependent cities and the oil-rich Gulf area cities (Peter Newman, 2016); (Manaugh && El-Geneidy, 2011), which is an indication that future transportation will revolve around the train. All of these patterns were predicated on the faster-than-average traffic in Asian cities, as well as re-urbanization tendencies focused on TODs, which are increasingly the locations for knowledge economy employment and improved accessibility (Kenworthy, 2015).

With tremendous enthusiasm, Chinese and Indian cities adopted this trend, planning or constructing 82 Metro rail projects in Chinese cities during the previous ten years and 51 in Indian towns. "In India, metro urban rail is currently operational in nine cities with a built-out 379km network; another eight cities are building out 277km of the metro rail network; and a further 20 cities have rail transit under initiation with any city over a million now eligible for Federal assistance(Anon., 2017)." Although all initiatives have political support, they are unable to be built since they need so much finance.

## Literature Review

A comprehensive literature evaluation was done because the research for this publication is based on many case studies. Three case studies have been studied extensively in this report. Case study of Shanghai, China, Case study of Recife metro, Brazil and case study of Delhi, India have been presented below in the tabular column. Based on the extensive study on case studies, a similar kind of case study for evaluation of Impact of Namma Metro on Bengaluru city has been attempted in this paper.

## Methodology

In a well-structured format, closed-ended questions have been used for this investigation. When creating the questionnaire, the questions needed to assess a socioeconomic impact were taken into account. The questionnaire was sent out to various Metro riders and people living in Metro corridors. The questionnaire was also disseminated on social media sites via a Google forms link. The needed sample size to examine the impact of Namma Metro on Bangalore city was calculated using a finite population sampling approach. The information obtained from the survey and Google forms was analysed to determine how the Metro Phase 1has affected the neighbourhood. Phase 1 of Purple line Namma Metro was inaugurated in the year 2011 and a study after decade will throw light in the impacts and influences caused by the infra development influx to the city.

Sample size to analyse social and economic impact of Namma Metro on Bangalore city were calculated by using Krejcie and Morgan sampling method(Morgan, 1970).

$$n = \frac{z^2 * N * \sigma_p^2}{(N-1)e^2 + z^2 \sigma_p^2}$$

Where, n= Sample size, N= Population of city, Z=Critical value, σ=Standard deviation
e=Acceptable error

The confidence level used for determination of sample size is 95% with critical value Z as 1.96. Population of Bangalore city is around seventy-three lakhs as per 2011 census, while acceptable error (e) is assumed to be 0.05 and the standard deviation to be 0.5. So, the sample size obtained by considering above constants is 384.

| | | SHANGHAI, CHINA-CASE STUDY (Pan & Zhang, 2008) | | |
|---|---|---|---|---|
| **Sl. No** | **Details of the projects** | **Parameters & observation** | | **Result** |
| 1 | Rail Transit Impacts on Land Use: Evidence from Shanghai, China, Haixiao Pan and Ming Zhang | Functional composition was conducted with an inner buffer of 200 m and doughnut shaped buffer between 200 m & 500 m from the station | | In more accessible regions close to train stations, more capital-intensive land uses and higher development intensity occur. |
| | Description: Changes in land use associated with rail transit, All the 3 metro lines were considered | **Seven lands use categories**—residential, office usage, commercial, other public facilities, industrial and warehousing, green space, and transportation—are used to group together data on land use for each buffer zone. | The inner buffer is used for residential purposes to a lesser extent than the outside buffer. On the other hand, the inner buffer for all three metro lines contains a greater proportion of commercial land use than the outer buffer. In comparison to Lines 1 and 2, Line 3 station locations have a substantially higher percentage of land used for warehouse or industrial purposes. | Hedonic price modeling shows that the transit proximity premium amounts to approximately 152 yuan/m2 for every 100 m closer to a metro station. |
| | | **Development Intensity:** The values of FAR were calculated for allotments within a 500-m radius of the stations. The intensity of development was then classified as low, medium, or high. | Around Line 1 and Line 2 metro stations, the inner buffer has been developed more thoroughly than the outside buffer. Contrarily, on Line 3, the inner buffer was even less developed than the outside buffer. | In Shanghai, rail transit is a magnet that attracts new development or redevelopment to areas that the system covers. |
| | Property Value Impacts: | **Structural characteristics,** which include total floor area, lot &size, single- or mixed-use unit, age, finished interior or structure onlyand site density | According to the findings, the sales prices of residential units are affected differently by the distance to the station as well as other factors related to location and neighbourhood. | |
| | | **Neighborhood characteristics**, Local traits mostly consist of neighbourhood amenities, such as the availability of parks, stores or other retail establishments, schools, hospitals, and sports facilities. | | |

| | **RECIFE, BRAZIL - CASE STUDY**(Andrade & Maia, 2009) | | | |
|---|---|---|---|---|
| **Sl. No** | **Details of the project** | **Parameters & observation** | | **Result** |
| 2 | **Paper Title & Author:** The Recife Metro – the Impact on Urban Development after 20 years, Mauricio Oliveira de Andrade & Maria Leonor Alves Maia | The Hedonistic Pricing Technique was used to determine how the price of any urban land responds to the recent metro influx to the city. When all other qualities or components that can explain the worth of an item are kept constant, the hedonistic or implicit price of an attribute is equal to its volatility. | | The final result of the study establishes a formula which derives and explains about the land value of a place or neighbourhood to various factors and closeness to
1. City centre,
2. Variable according to existing land use
3. Transport hub or intermodal hub
4. Size of the site
5, Security concerns due to neighboring environment
4. Any MRTS station
5. Legal issues of the lay out with a correcting variable
6. Development services and other infrastructure
7. A variable according to the development indexes and
8. Occupancy index as per the density

The study establishes a scientific way of quantifying the influence |
| | **Description of the study and study area:** To research the link between modes of transportation and land use in the Recife area of Brazil as well as the impact of metropolitan railway lines on urban growth. Brazil's Recife Metropolitan Railway With a population of 3 million, Recife is one of the major metropolitan areas in northeastern Brazil. With a population of 4,054,866, Recife is the state capital and largest city of Pernambuco in the northeastern region of South America. Additionally, it is the largest urban region in the Northeast and North Regions. The city proper's populace in 2020 was 1,653,461. The SEI (Integrated Structural System), which also has 49 bus routes and seven terminals, | **Location-related characteristics:** The separation between the main area, the closest metro stop, and the closest significant transit centre. | Had little impact, according to a comparison of the value of land in the areas along several transit tracks. The model shows a concentration of the increase in value around metro stations, although this increase is not greater than in other regions of the city. | |
| | | **Neighborhood-specific characteristics:** Infrastructure, the primary use of the land, the HDI, the median household income, population density, and the number of homicides per 10,000 people are a few examples. | The rise in the value is likely the use of metro among the residents of the informal settlements that almost half the population and found rise of 28% users in 2004 to 2.7% users in 1984. | |
| | | **Physical Qualities:** Size, dominating breadth, prevailing terrain, and geometrical or practical characteristics of the transit route. | However, conflicts between economic, political, and social pressures also dictate the type and extent of modifications that must be made, as well as their impacts. Clearly, investing in transportation is crucial to helping cities attain a balanced urban layout. | |

| | | | |
|---|---|---|---|
| incorporates the Recife Metro as part of its network. | | | of the transport mode on property prices. |

| \multicolumn{5}{|c|}{**DELHI, INDIA - CASE STUDY** (CentralTransportation Plannlng and Environment Division , 2007)} |
|---|---|---|---|---|
| **Sl. No** | **Details of the project** | \multicolumn{2}{c|}{**Parameters & observation**} | **Result** |
| 3 | **Paper Title & Author:** Contribution of Delhi Metro Rail Corporation (DMRC) towards Betterment of Delhi's Environment<br><br>**Description of the study and study area:** To conduct the Environmental Impact Assessment (EIA) of Delhi metro, India. Delhi is a union territory and city in India, home to New Delhi, the nation's capital. It is surrounded by Haryana on three sides, while Uttar Pradesh borders it on the east. 1,484 square kilometers make to the NCT (573 sq mi). The city proper of Delhi, India's second biggest after Mumbai, with a population of over 11 million people as per the 2011 census. The construction of rail-based metro systems in 11 cities throughout the nation was made possible by the Delhi Metro, India's first modern metro rail project. | \multicolumn{2}{l|}{A checklist was created based on the data collected and was presented} | The Delhi Metro will contribute to reducing the pollution load brought on by car traffic, which is rapidly rising. Therefore, there will be huge reductions in both fuel usage (about 7 million liters of diesel, 31.5 million liters of gasoline, and 16.68 million kilo grammes of CNG, as specified in the text) and foreign exchange losses (a staggering Rs. 1722 million). |
| | | **Project Location Impact** | 1. Had negative impact on rehabilitation and resettlement, Change of land use and Ecology, Drainage, and utility problems.<br>2. Had no impact on historical and cultural monuments | |
| | | **Project design Construction Impact** | 1. Had no impact on platform inlets and outlets, ventilation and lighting and risks due to earthquake.<br>2. Had negative impact on railway station refuse | |
| | | **Project construction impact** | 1. Had a detrimental effect on soil erosion, pollution, and health risks at construction sites, as well as risk to existing buildings and traffic diversion. | |
| | | **Project operation impact** | 1. Had negative impact on oil pollution, noise and vibration and water demands | |
| | | **Environmental impact** | 1. Had a favorable effect on increased employment possibilities, economic growth, service and safety, traffic congestion reduction, and fuel efficiency. less smog in the air. savings in the construction of roads | |

## Study Area

Bangalore, referred to as Bengaluru, is the capital of the Indian state of Karnataka, which is situated in the country's south at 12.97° N and 77.56° E. Bangalore is situated in the centre of the Deccan Plateau's Mysuru Plateau. The city, which was once planned as a garden city, has changed over time to become India's industrial and software powerhouse. As domestic and international businesses were established during this time, Bangalore's IT sector expanded. The epicenter of economic activity is Bangalore. Bangalore's population has increased dramatically because of rapid urbanisation and industrialization and is projected to reach 14.1 million by 2021. It is India's third most populous city and fifth-most populous urban agglomeration by metropolitan population, as measured by the 2011 Census, with a metro population of close to 8.5 million.

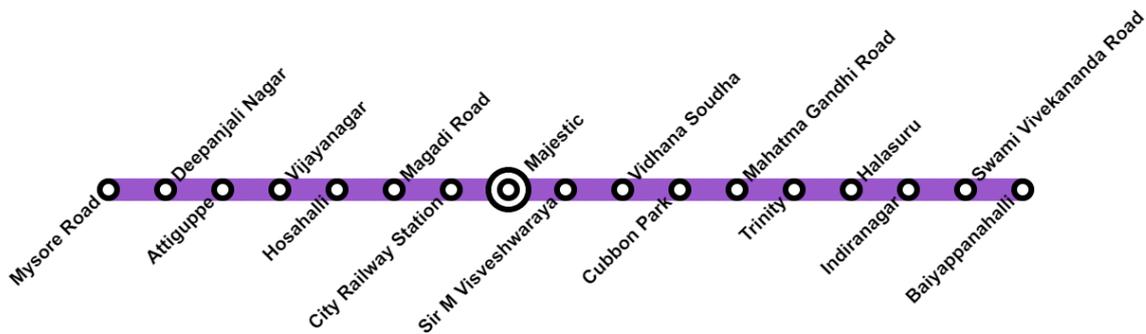

**Figure 1 : Bengaluru metro Rail Phase-1**

## Data Collection and Data Analysis

In this paper, Part of Purple Line Phase-1, was considered for the evaluation impact of Socio-economic characteristics of Namma Metro on Bengaluru city. In one of the studies conducted for a Metro station along the purple line showed that, it has helped in decongestion of the traffic and thus aiding in sustainable development of the city (R.Hemasree, n.d.). One more study along the purple line, to understand the relevance of the location also concluded that Metro rail systems helps in promoting the Non -motorized transport and also reduction in traffic congestion, pollution in a positivity note but also suggesting for a better planning of parking and other facilities for better use of Metro rail system.(Hemasree & Subramanian, 2019). The study was conducted along the corridor Byappanahalli to Majestic of purple line i.e., East-West corridor. In this paper, the survey was conducted for the people along E-W corridor and hence the ¼ sample size is considered for sampling which are around 96. The study was conducted on the metro riders and also on the people living in Metro corridors. From the questionnaire 120 responses were obtained and after examination could take 97 responses from Metro riders and 69 responses from people living in Metro corridors were found eligible in terms of responses for all the questions.

**The responses from the metro riders:**

The commuters using Metro were provided with the questionnaire and the response to those questions are shown in the graph below

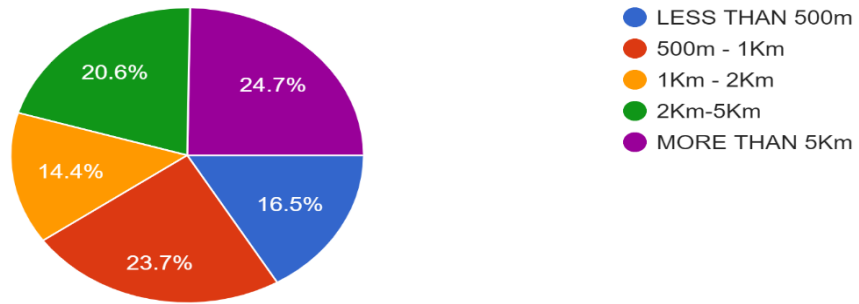

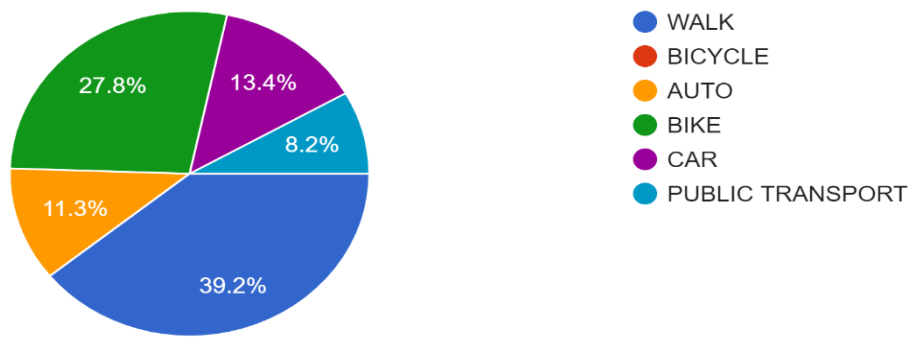

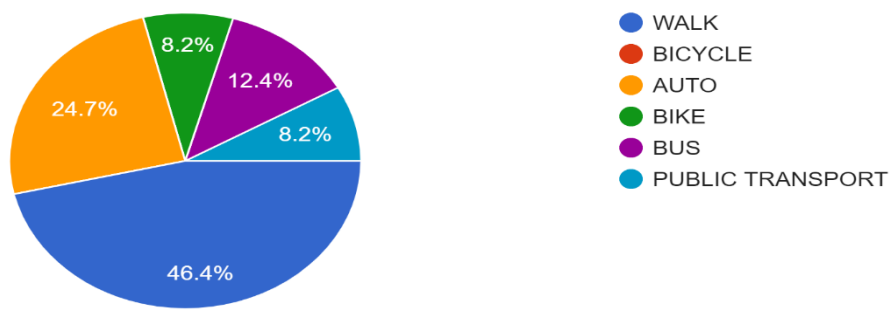

## HOW OFTEN DO YOU USE THE NAMMA METRO?
97 responses

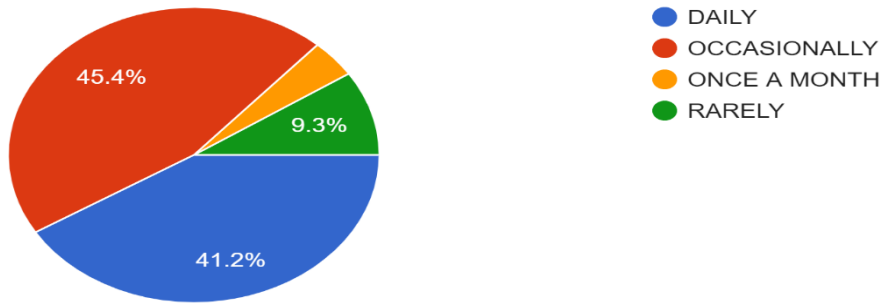

- DAILY — 41.2%
- OCCASIONALLY — 45.4%
- ONCE A MONTH
- RARELY — 9.3%

## WHY DO YOU CHOOSE METRO OVER OTHER MEANS OF TRANSPORT?
97 responses

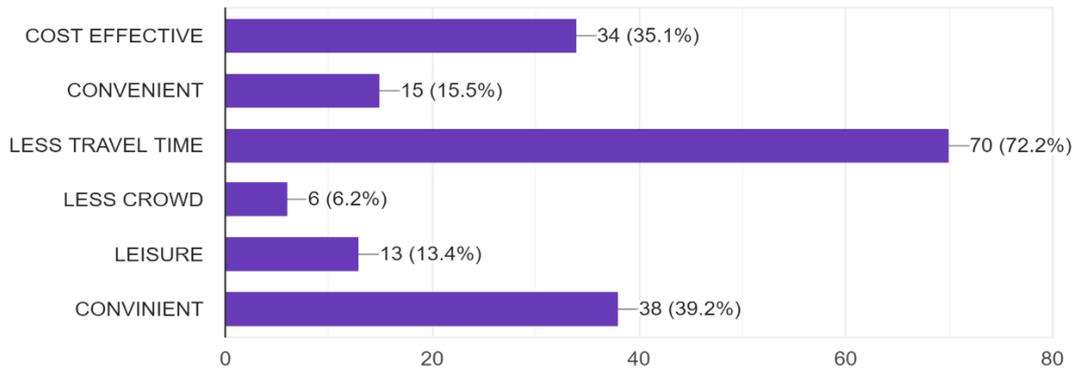

- COST EFFECTIVE — 34 (35.1%)
- CONVENIENT — 15 (15.5%)
- LESS TRAVEL TIME — 70 (72.2%)
- LESS CROWD — 6 (6.2%)
- LEISURE — 13 (13.4%)
- CONVINIENT — 38 (39.2%)

## Do you think that the Namma Metro project has reduced the traffic problems?
97 responses

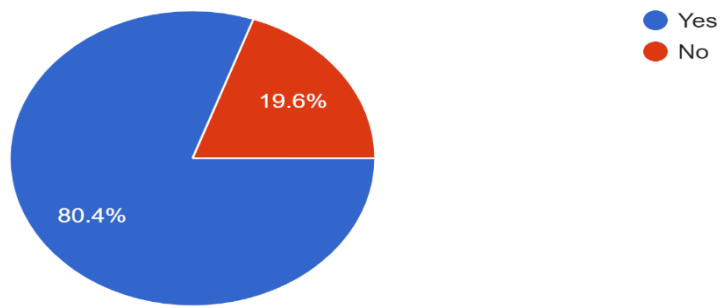

- Yes — 80.4%
- No — 19.6%

Do you think the Namma Metro height spoils the beauty and affects visual aesthetics and permeability?
97 responses

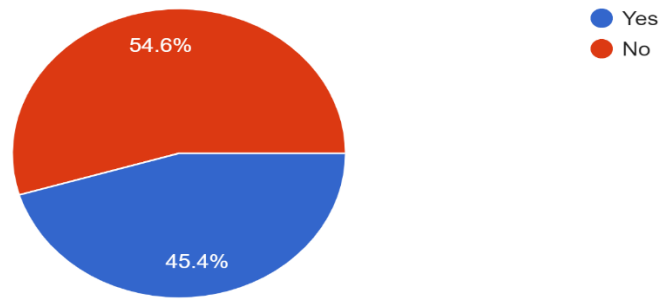

- Yes
- No

54.6%
45.4%

Did the Namma Metro project impact the surrounding city environment?
97 responses

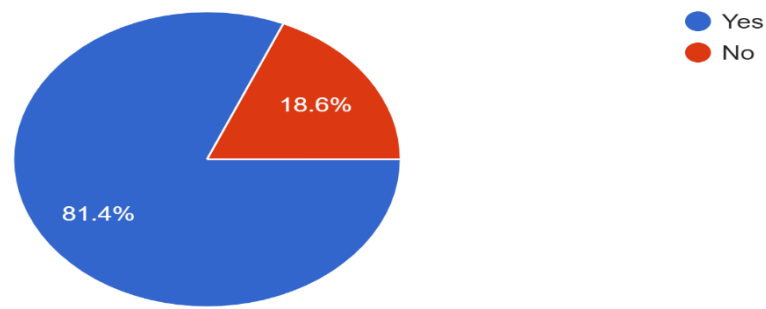

- Yes
- No

81.4%
18.6%

**Figure 2 : Responses from the commuters of Metro.**

From the above responses we can observe that, majority of the Namma Metro users up to 5Km reach the station by walking, cycling, use two wheelers or a public transportthus reducing the traffic congestion and also increasing the non-motorized transport means for reaching the Metro station. More than 45 % of the people use Metro for their work trip as it saves time. Majority of the people (more than 80%) believe that Metro has reduced the traffic problems and impacts the surrounding environment. Commuters using Metroas their major transport mode are happy with the services.

**The responses from the people living in metro corridors:**

People living in the Metro corridors were asked to rank the importance of the metro's influence during the study.Questionnaire was prepared to understand an impact on the surrounding environment due to Namma Metro construction. The questionnaire included the aspects such as impact on surrounding environment, losing of buildings, dust, air pollution,and visual aesthetics of locality, reduction in traffic and transportation problems. The responses obtained from the survey are as shown below in Figure 3.

Has the Namma Metro project construction created an impact on the surrounding environment?
69 responses

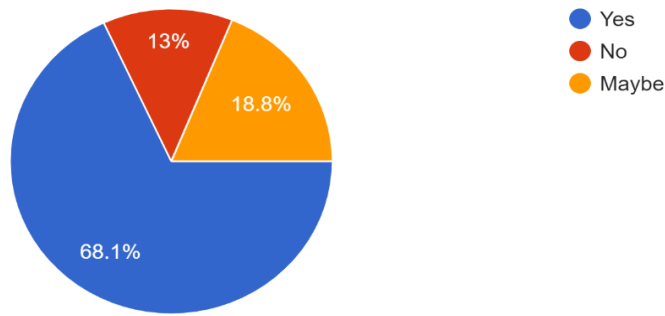

Did the Namma Metro project construction create issues of dust, sound pollution and traffic jams, etc.
69 responses

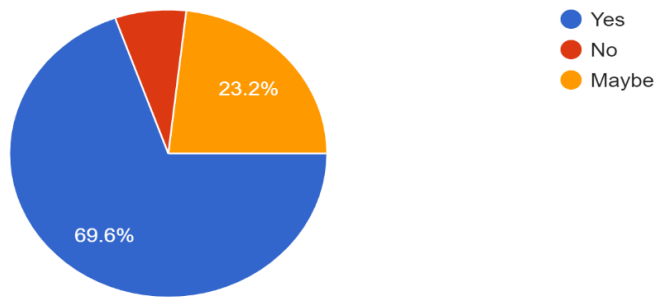

Did you lose any important structure in your locality due to Namma Metro?
69 responses

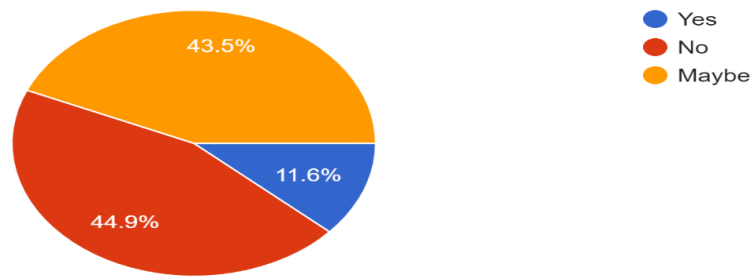

Do you think the Namma Metro height spoils the beauty and affects visual aesthetics of the locality?
69 responses

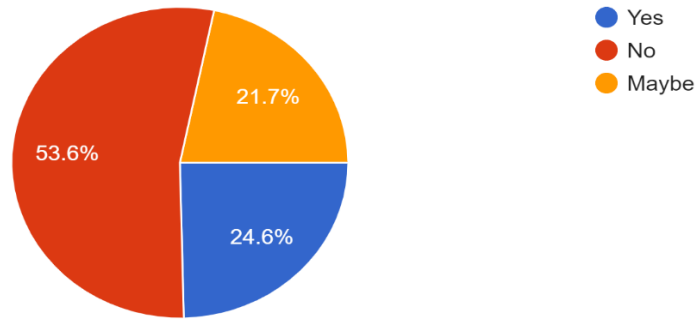

Do you think that the Namma Metro project has reduced the air and noise pollution?
69 responses

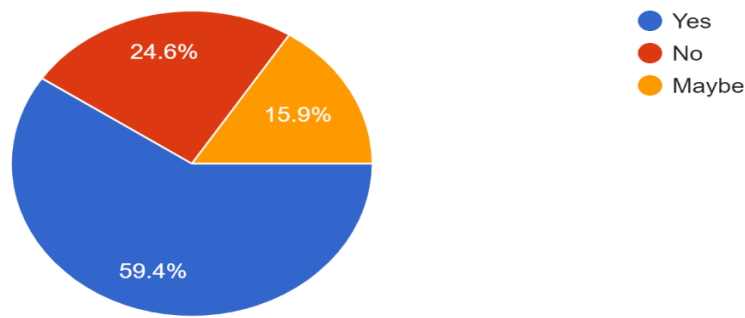

Do you think that the Namma Metro project has reduced the traffic and transportation problems
69 responses

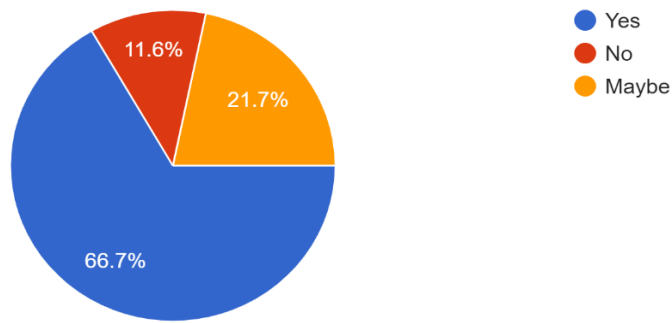

**Figure 3 : The graph of Reponses for various questions during survey.**

From the above responses, it can be observed that there is a positive impact on various factors such as reduction in pollution, reduced traffic and transport problems, and visual aesthetics apart from the negative impacts such as vibration, dust and air pollution during construction. Some of them also spoke about of Iconic buildings they lost in the process. More explorations are needed to analyse the other segments of the city like sense of place, Imagibility, accessibility and the effect of TOD on the city of Bengaluru.

**Summary and Findings:**

**(i) The responses from the metro riders:**

The Metro riders are found to be of around 56% female and 44% male ,around 70% of them aged from 18 to 50 years of age, around 74% of them graduate and post graduate in education, Students around 23% and private employed 35%, middle and upper middle class, and majority of them almost 60% travelling for work and education. They seem to adhere to this MRTS for its convenience and time saving results and they believe that Namma metro has solved the traffic problems of the city. More than 80% of them are aware of the influences made by the Metro on the city. They have overruled the negative impacts caused by Metro on the aesthetics of the city and believe that Metro solves the traffic problems and has positive effects on the city environment.

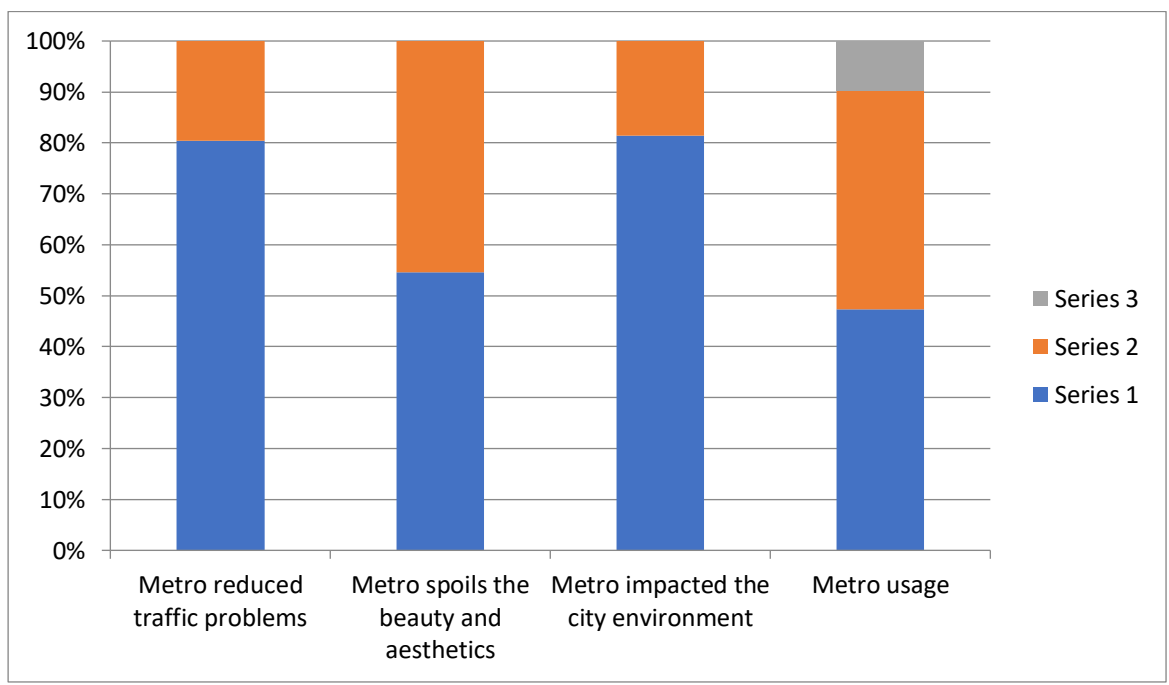

**Figure 4: The graph representing summary of responses for the questions about the influences and impacts in the minds of Namma Metro riders.**

**(ii) The responses from the people residing on the metro corridors:**

Metro riders are found to be of around 48% female and 52% male ,around 65% of them aged from 18 to 50 years of age, around 79% of them graduate and post graduate in education, Students around 18% and private employed 45%, middle and upper middle class, and majority of them almost 50% travelling for work and education. Almost 50% of them use the metro facility everyday and 40% use the facility occasionally. Almost 90% of them had been affected by Metro construction and almost 70% of them received compensation for their property. 45% of them say they lost some iconic structure which was giving them visual legibility and 69% believe that Metro has impacted their environment.54% of them feel that Metro has affected the aesthetics of their environment .After all the long list of negative impacts caused by Namma Metro around 60% of them accept that metro has caused positive impacts on the environment and 67% of them accept that Metro has solved the traffic problems of the city. Around 50% of them even accept that they are very satisfied with the services and operation of Namma Metro.

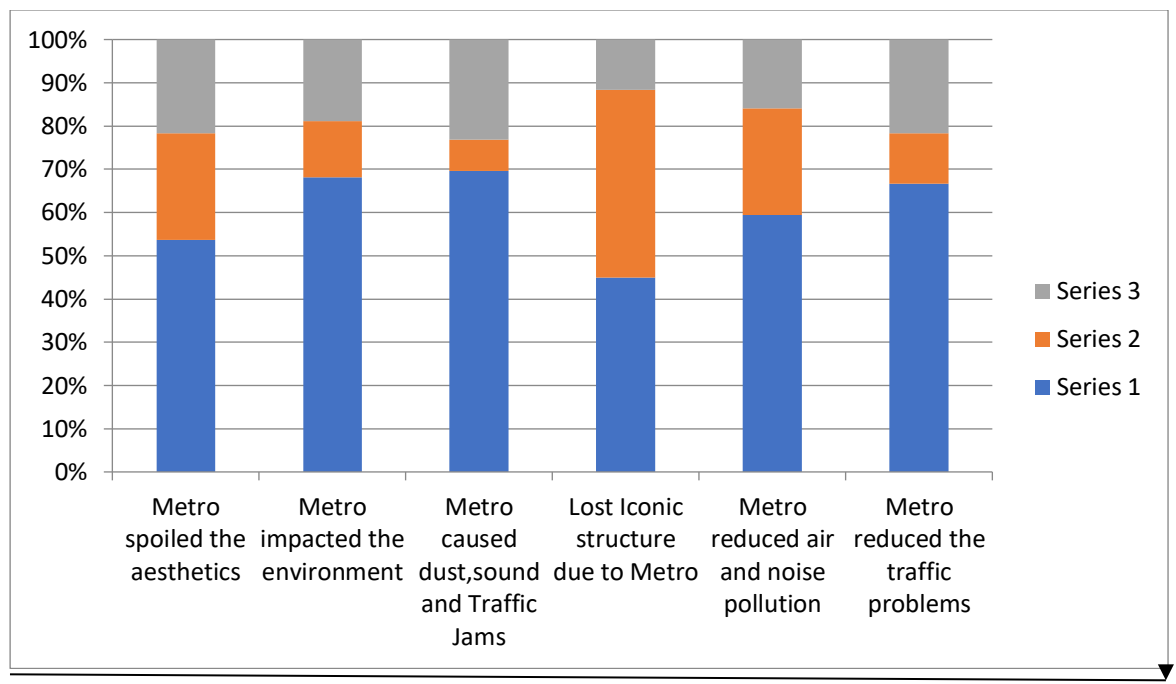

**Figure 5: The graph representing summary of responses for the questions about the influences and impacts in the minds of people residing on the Metro corridors.**

**Conclusion:**

The Namma Metro in Bengaluru has several advantages, including a decrease in air pollution, time savings for passengers, a decrease in accidents, a decrease in traffic congestion, and fuel savings, all of which influence the socioeconomic position of commuters on the purple line corridor. The various parameters examined in the survey and their outcome can be summarized as follows (Table 1).

Table 1 : Parameters and Observations from the Study area

| Parameters & observation ||
|---|---|
| **Location-related characteristics:** Distance from the metro station, accessibility to the metro station, and convenience of using Metro. | Had positive impact, as most of the people are using Metro for their work trip, which reduces time of travel and convenient for travel. |
| **Neighborhood-specific characteristics**, such as the reduced traffic problems, visual aesthetics of the area and impact to the surrounding environment. | The metro commuters have a positive attitude towards the Metro as the survey says it reduces the traffic congestion problem and does not impact the aesthetics of the locality. It also promotes non-motorized transport. |
| **Environmental Characteristics:** Traffic and transportation problems, air and noise pollution, aesthetics of the locality, impact during construction of metro | People living in Metro corridors during survey also gave positive feedback about the Metro services in terms of reduction of traffic jams, reduction in air, noise, and dust pollution. Only during construction, they experienced some inconvenience due dust and diversions along the construction road. |

# QUESTIONAIRE A &B

## RESEARCH SURVEY-NAMMA METRO-BENGALURU
## FORM -A-COMMUTERS IN METRO

Ar.R.Hemasree, Research Scholar, Guided by Dr.C.V.Subramanian Department of Architecture, PMIST.

**Name**: ______________________________________

**Age:** ☐Under17  ☐18-30  ☐ 31-59  ☐ Above 59

**Gender:** ☐Male      ☐Female

**Education:**   ☐ 10th ☐ 10+2  ☐Diploma ☐ Graduate ☐Post Graduate ☐OtheRs.

**Occupation**: ☐ Student   ☐ Private Employed  ☐Government Employed  ☐Self Employed

☐Retired  ☐Unemployed

**Monthly Income**: ☐Up to Rs.30,000  ☐Rs.30,000-60,000  ☐ More than Rs.60,000

**Have you ever travelled by Namma Metro?**

☐Yes   ☐No

**Place of Residence:** _________________

**Place of destination**: _________________

**What is the usual timing using the Namma Metro services?**

☐ Morning  ☐Afternoon   ☐Evening   ☐Night

**What is the purpose of the trip?**

☐ Work  ☐Education   ☐Leisure   ☐Personal   ☐ Other

**What is the distance between your home and the Namma Metro station?**

☐ < 500 m ☐500 m- 1 km   ☐1 km - 2 km   ☐2 km – 5 km   ☐ >5km

**What is the mode of transportation you take to reach the Namma Metro service?**

☐ Walk  ☐Cycle   ☐Auto  ☐Bike   ☐ Car   ☐ Public transport

**What is the mode of transportation after reaching the destination?**

☐ Walk  ☐Cycle   ☐Auto  ☐Bike   ☐ Car   ☐ Public transport

**How often do you avail the Namma Metro services?**

☐ Daily  ☐ Occasionally  ☐ Weekly once  ☐ More than once weekly

**What is your expenditure on using Namma Metro services?**

☐ < Rs.10   ☐ Rs.11-30   ☐ Rs.30-50   ☐ Rs.50-100   ☐ > Rs.50

**How satisfied are you with the Namma Metro services?**

☐ Extremely Satisfied  ☐ Satisfied  ☐ Neutral  ☐ Dissatisfied  ☐ Extremely Dissatisfied

**Why do you choose Namma Metro over other services?**

☐ Cost effective  ☐ Convenient  ☐ Less time factor  ☐ Less crowded  ☐ Leisure purpose

**Did you hear of proposed Namma Metro construction in this route as major project?**

☐ Yes    ☐ No

**Do you think Namma Metro would be the best public transportation facility compared to other transit services?**

☐ Yes    ☐ No

**Do you think that the Namma Metro project has reduced the traffic problems?**

☐ Yes    ☐ No

**Do you think the Namma Metro height spoils the beauty and affects visual aesthetics and permeability?**

☐ Yes    ☐ No

**Did the Namma Metro project impact the surrounding city environment?**

☐ Yes    ☐ No

**RESEARCH SURVEY-NAMMA METRO-BENGALURU**

**FORM -B-PEOPLE LIVING ON MERO CORRIDORS**

Ar.R.Hemasree, Research Scholar, Guided by Dr.C.V.Subramanian Department of Architecture, PMIST.

**Name**: ________________________________________

**Age:** ☐ Under17    ☐ 18-25   ☐ 26-59   ☐ Above 59

**Gender:**   ☐ Male           ☐ Female

**Education:** ☐ 10th ☐ 10+2 ☐ Diploma ☐ Graduate ☐ Post Graduate ☐ OtheRs.

**Occupation:** ☐ Student ☐ Private Employed ☐ Government Employed ☐ Self Employed ☐ Retired ☐ Unemployed

**Monthly Income:** ☐ Up to Rs.30,000 ☐ Rs.30,000-60,000 ☐ More than Rs.60,000

**Have you ever travelled by Namma Metro?** ☐ Yes ☐ No

**Place of Residence:** ________________

**Place of Destination:** ________________

**How many years you have been residing in this locality?** ________________

**How often do you avail the Namma Metro services?**

☐ Daily ☐ Occasionally ☐ Weekly once ☐ More than once weekly

**What is the mode of transportation you take to reach the Namma Metro service?**

☐ Walk ☐ Cycle ☐ Auto ☐ Bike ☐ Car ☐ Public transport

**Did you hear of proposed Namma Metro construction in this route as major project?**

☐ Yes ☐ No

**Has your property been affected with the Namma Metro project construction?**

☐ Yes ☐ No

**If your property was affected did you receive any compensation?**

☐ Yes ☐ No

**Did the Namma Metro project construction create issues of dust, sound pollution and traffic jams, etc.**

☐ Yes ☐ No

**Did you lose any important structure in your locality due to Namma Metro?**

☐ Yes ☐ No

**Has the Namma Metro project construction created an impact on the surrounding environment?**

☐ Yes ☐ No

Reason_________________________________________________

**Do you think the Namma Metro height spoils the beauty and affects visual aesthetics of the locality?**

☐Yes    ☐No

**What has been the effect of the Namma Metro construction on the business sector?**
__________________________________________________

**Do you think that the Namma Metro project has reduced the air and noise pollution?**

☐Yes   ☐No

**Do you think that the Namma Metro project has reduced the traffic and transportation problems?**

☐Yes   ☐No

**Are you satisfied with the Namma Metro project construction and its services?**

☐ Extremely satisfied ☐ Satisfied ☐ Neutral   ☐ Dissatisfied  ☐ Extremely dissatisfied